\documentclass[textures,12pt]{article}
\usepackage{epsfig}
\def\beq{\begin{equation}}
\def\eeq{\end{equation}}
\def\ben{\begin{eqnarray}}
\def\een{\end{eqnarray}}
\def\bea{\begin{array}}
\def\eea{\end{array}}

\begin{document}

\baselineskip=20pt

%------------------------------------------------------------------------
%   BEGIN THE TITLEPAGE 
%------------------------------------------------------------------------

\begin{flushright}
Prairie View A \& M, HEP-7-97\\
September 1997 \\
Los Alamos National Lab bulletin board: hep-ph/9710485
%({\bf Draft})

\end{flushright}

\vskip.15in

\begin{center}
{\Large\bf Exploring inconsistencies with the 
Cabibbo-Kobayashi-Maskawa model}

\vskip .3in

{\bf Dan-Di Wu}\footnote{ $~$E-mail:  wu@hp75.pvamu.edu or 
danwu@physics.rice.edu }
\vskip.6in

{\sl HEP, Prairie View A\&M University, Prairie View, TX 77446-0355, USA}
\end{center}
\vskip.36in

{\sl A small upper-bound is obtained for the CP phase $\delta$ in a CKM matrix,
 $|\sin\delta| < 0.44$ 
based on CP  independent data on $|V_{ub}/V_{cb}|$ alone.  
Potential inconsistencies with the
CKM model in existing data which need a theory beyond the standard model  
to overcome emerge in the analysis using this CKM matrix.
In addition,  all CP asymmetry measurables for
the $B_d-\bar B_d$ system are expressed solely by
the CP phases in the decay amplitudes. Remarkably 
the  CP asymmetry of the bench mark process 
$B_d\rightarrow \psi K_S$ equals  sin$2\delta$.
This will soon be measured by BaBar and Belle to 
confront  the above upper-bound. 
    
 }
\newpage
The Cabibbo-Kobayashi-Maskawa (CKM) model\cite{CKM} 
is the best available frame work for the study of weak interactions involving
quarks and for the study of CP violation. The CKM model is also an 
indispensable part of 
the standard model (SM). An approach in the effort for finding
 new physics beyond CKM is  
to find data which are inconsistent with the CKM model. In doing so, 
better and more accurate data are of course crucial. However the success 
is also very much dependent on the specific CKM matrix one is using in the 
analysis. For example, 
the matrix elements $V_{ud}$ and $V_{us}$ are already 
measured to high accuracies.
The two exact values seem not enough to provide a test of
their consistency with the CKM
model if one uses the original form of the Kobayashi-Maskawa matrix because 
functions of other angles are involved, in addition to the Cabibbo angle.
With the Wolfenstein matrix\cite{Wol},
the two matrix elements are essentially different functions of
 one and the same parameter 
$\lambda$. Consequently, the two values provide a crucial consistency check 
with the CKM model\footnote{With the Wolfenstein matrix, $V_{ud}$ and $V_{us}$
will not be functions of parameters other than $\lambda$ up to the order of $
\lambda^5$, and the coefficient of the $\lambda^6$ term is of order 1, 
see later.}. 

\vskip.1in

 Several forms of the CKM 
matrix\cite{CKM}\cite{Wol}\cite{Chau}\cite{databook}\cite{Chen}\cite{Xing97}
have appeared in the literature. 
(For simplicity, different forms will be called 
different CKM matrices in this note.)
However, popular CKM matrices\cite{databook} are only convenient
for $K^0-\bar K^0$ studies, not for B-physics studies 
(see later). 
Indeed the phase convention dependent 
mixing parameter $\epsilon_{_K}$ is very 
small in the
popular CKM matrices; 
however  $\epsilon_{_{B_d}}$ in these matrices is not small 
enough for any convenience.
 In this note 
we  will recommend two new CKM matrices, based on a previous work 
of the author with Chen\cite{Chen}. 
One is convenient for 
$K^0, B_d$ and $D^0$ and the other is convenient for $B_d, B_s$ and $ D^0$.
The values of the parameters of such matrices will  be briefly 
discussed\footnote{For a recent study on CKM parameters, see Ref\cite{Nir}
}. A 5.2 $\sigma$ difference is there in the
$\lambda$ value measured by $d$-quark decays and  by $K$ decays, which, 
if persists, will be a signal of inconsistency of experimental data with the 
CKM matrix.
A  small
upper-bound for sin$\delta$ is obtained from the CP independent 
data on $|V_{ub}/V_{cb}|$, where $\delta$ is the 
only CP phase  in these matrices.
This bound  may also jeopardize the standard model, 
if the already small value becomes 
even smaller along with more accurate data. This bound  or its 
equivalent bound with similarly clear physical meaning 
does not appear if one works with other CKM matrices.
A big advantage of using these matrices is that the CP phases in the 
mixing mass  and mixing width 
 are extremely small for the neutral particles involving
the b quark. Consequently, all CP
measurables are exclusively expressed by the CP phases in the decay amplitudes.
This will greatly simplify  data analysis in the near future. 
One of the remarkable results of this study 
is that the  asymmetry of the
bench mark process $B_d\rightarrow \psi K_S$\cite{Sanda}
will be exactly sin$2\delta$ in Eq (17) (the master equation).   
This asymmetry is expected to be 
measured in the first phase of BaBar and Belle experiments.
One will soon be able to confront the measured $\delta$
 with the above upper-bound.
\vskip.08in
The note is organized as the following. In the beginning, 
academic considerations 
are presented to evaluate the characters of different CKM matrices,
in terms of whether they provide a small mixing parameter $\epsilon$
for a specific mixing system. A reader   may skip this part 
if he does not want to go into academic details.
The advantage of having extremely small $\epsilon$ for the B-physics 
is clearly expressed by Eq (8). The CKM matrix which is convenient for 
$K$, $B_d$ and $D$ studies is presented in (9) and its simplification, in (10).
In discussion of the values of the parameters in the CKM matrix (10), 
an upper-bound for the CP phase $|sin\delta|$
is obtained in Eq (15).  The implications of Eq (8) are further explored,  
with Eq (17) presented. In Eq (20) we present 
a CKM matrix which is convenient for heavy flavor studies.
\vskip.1in
A phase convention dependent quantity $\epsilon$ has played an important role in 
$P-\bar P$ (Here $P $ stands for a neutral pseudo-scalar particle) mixing 
and CP violation studies. $\epsilon$ is defined as a parameter which appears in 
the definition of mass eigenstates\cite{TWu}, assuming CPT conservation.
\beq
|P_{\mp}\rangle=\frac{(1+\epsilon)|P\rangle\pm (1-\epsilon)|\bar P\rangle}
{(1+|\epsilon|^2)^{1/2}}.
\eeq
%1
 $\epsilon$ is commonly called the mixing parameter, although
it does not bear an immediate physical meaning: 
it is not measurable, for example. What is physically important for $\epsilon$
is the value of $|(1+\epsilon)/(1-\epsilon)|$. Obviously, this value does not
change when, for instance, $\epsilon\rightarrow \epsilon^*$, and  $\epsilon
\rightarrow \frac{1}{\epsilon}$.
\vskip.08in
 An essential and related quantity,
$\sigma$, 
called the overlap,  is defined as
\beq
\sigma=\frac{1}{2}\langle P_+|P_-\rangle.
\eeq
%2
$\sigma$ is a phase convention independent
quantity\cite{Wu81}
\beq
\sigma=\frac{{\rm Im} M^*_{12}\Gamma_{12}}{4|M_{12}|^2+|\Gamma_{12}|^2}
\eeq
%3
which is a real number for all mixing systems. For the $B_d$ and $B_s$
systems, $\sigma=\frac{1}{4}$Im$(\frac{\Gamma_{12}}{M_{12}})$.
It is also directly measurable 
 through the measurement of the asymmetry in the
leptonic  decay channels\cite{Pais} $P_+ \rightarrow l^\pm + x (\bar x)$. 
 For the $B_d$ system $\sigma$ is approximately (to calculate
$M_{12}$ and $\Gamma_{12}$, see Ref\cite{Inami}\cite{Hagelin})
\beq
\sigma_{_{B_d}} \simeq\pi \frac{m_c^2}{m_t^2}\sin\delta\le 2.2\times 10^{-4}.
\eeq
Values of $\sigma$ for other mixing systems are given in Table 1.
%4
\vskip.08in
It has been shown\cite{Kabir}
 that in a complex plane, the possible values of $\epsilon$ makes a
 loop whose  diameter is exactly $|\sigma - \frac{1}{\sigma}|$. 
This loop is big, especially when $\sigma$ is as small as that in Eq (4). 
The extremes 
of $|\epsilon|$ appears when $\epsilon$ is real and is exactly
 $\epsilon=\sigma$ (the minimal magnitude) 
and $\epsilon=\frac{1}{\sigma}$ 
(the maximal magnitude). 
This can also be seen  from
the complex expression for $\epsilon$,
\beq
\epsilon  =\frac{i{\rm Im} M_{12}+{\rm Im}\Gamma_{12}/2}{
{\rm Re} M_{12}-\Delta m/2-(i/2) {\rm Re} \Gamma_{12}+i\Delta\gamma/4},
\eeq
%5
where
\ben
\bea{ccc}
\Delta m &=&m_- - m_+\\
\Delta \gamma&=&\gamma_--\gamma_+
\eea\een
%6
and $m_\pm$ are the masses of $P_\pm$, and $\gamma_\pm$ are the widths of 
$P_\pm$ respectively. These definitions seem to be complete in defining a 
definite $\epsilon$. However,
in practice there are problems. First, $M_{12}$ 
is phase dependent,
so both Im$M_{12}$ and Re$M_{12}$ can change their values and signs when
the CKM convention is changed; similarly for  $\Gamma_{12}$. Second,
one really does not know whether $\Delta m$  is 
positive or negative; in other words, one does not know 
whether $P_+$ or $P_-$ is heavier
in a specific situation. One does not know the sign of $\Delta \gamma$ either.
\vskip.08in
One does know the relative sign of $\Delta m$
and $\Delta\gamma$ which is defined by\cite{Wu81}
\beq
4{\rm Re} M_{12}^*\Gamma_{12}=\Delta m\Delta \gamma.
\eeq
%7
Because of the smallness of $\sigma$ for all $P-\bar P$ systems, one has
$$|\Delta m\Delta \gamma|=4|M_{12}\Gamma_{12}|$$
to a very good accuracy.
Assuming $m_-$ is larger than $m_+$ (which of course needs 
justification\footnote{For the interested reader, see Ref\cite{Wu89}.}
),
then it is desired to find a CKM matrix 
with which one obtains a small Im$M_{12}$ as well as
a negative Re$M_{12}$. Such a matrix 
will make the parameter $\epsilon$ small because 2Re$M_{12}$ and $-\Delta m$
in the denominator will add together. They could cancel each other 
if only Im$M_{12}$ is small, but Re$M_{12}$ is positive. In that case, 
2Re$M_{12}$$-\Delta m$ would be zero (so would be 2Re$\Gamma_{12}-
\Delta \gamma$), which causes $\epsilon$ to 
explode.
\vskip.08in
In all popular CKM matrices\cite{databook}, $\epsilon_{_K}$ is close to 
its minimum, which brings some convenience in the $K_S-K_L$ studies.
However, with the same CKM matrices,
 such advantage is completely eluded in the $B$ physics studies where
promising and copious CP violation phenomena are going to be the center of
intensive studies. 
 Such convenience includes
avoiding talking about a big unknown mixing parameter $\epsilon$ and a 
correspondingly elusive ``big CP violation in the mixing in the SM"\footnote{
A phase convention independent definition of ``CP violation in the mixing" 
is recently emphasized by Nir.\cite{Nir}}.
In addition, the expression for CP asymmetry quantities is  
simplified\cite{Kabir}
\beq
\lambda_f=\frac{(1-\epsilon)}{(1+\epsilon)}\frac{\bar A_f}{A_f}
=\frac{\bar A_f}{A_f}
\eeq
%8
where $A_f(\bar A_f)$ is the decay amplitude of $B_d(\bar B_d)$ to a 
neutral final state f, which can be either self conjugated or not. 
In other words, in such CKM matrices, 
the CP asymmetries are all caused by 
CP phases in the decay amplitudes.
This formula is applicable to the $B_d-\bar B_d$ system, but not to the Kaon or 
D system, 
because $\sigma$ for these two latter systems
are compatible with their CP violation event rates $R$, (see Table I).
\vskip.08in 
Such a CKM matrix 
can be easily reached by  reparametrization of the CKM matrix\cite{CKM}.
Indeed, a convenient CKM matrix for the $K$ and $B_d
$ studies is recommended as the following based on the Wolfenstein
matrix\cite{Wol} and a work by Chen and Wu\cite{Chen}:\\
\ben
V_{BK}=\left(\bea{ccc}
1-\frac{1}{2}\lambda^2-\frac{1}{8}\lambda^4&\lambda
&A\lambda^3(e^{-i\delta}-\zeta+\frac{1}{2}\lambda^2e^{-i\delta})\\
&&\\
-\lambda+A^2\lambda^5(\zeta e^{-i\delta}-\frac{1}{2})&1-\frac{1}{2}
\lambda^2-(\frac{1}{8}+\frac{1}{2}A^2)\lambda^4
&A\lambda^2(e^{-i\delta}+\zeta\lambda^2+\frac{1}{2}\lambda^2 e^{-i\delta})\\
&&\\
A\zeta\lambda^3(1+\frac{1}{2}\lambda^2)&
-A\lambda^2e^{i\delta}(1+\frac{1}{2}\lambda^2)&
1-A^2\lambda^4\eea\right)
\een
%9
where all parameters $\lambda, A, \zeta$ and $\delta$ are real.
The unitarity of this CKM matrix is accurate  
to $\lambda^5$. 
 One finds 
J=$A^2\zeta\lambda^6\sin\delta$ for every quartet of the matrix.
A simplified version of this CKM matrix, which is accurate enough for most
practical purposes, is\\
\ben
V_{BK}=\left(\bea{ccc}
1-\frac{1}{2}\lambda^2&\lambda
&A\lambda^3(e^{-i\delta}-\zeta)\\
&&\\
-\lambda+A^2\zeta\lambda^5e^{-i\delta}&1-\frac{1}{2}
\lambda^2&A\lambda^2e^{-i\delta}\\
&&\\
A\zeta\lambda^3&-A\lambda^2e^{i\delta}&1\eea\right)
\een
%10
The approximation from  Eq (9) to (10), requires at least one of 
(cos$\delta - \zeta $) and
 sin$\delta$ to be of order 1.
The signs of $A$, $\lambda$, and $\zeta$ are adjustable by using ($-1$) 
to change the phase of a line or/and a column together with a redefinition
of $\delta$. It is therefore allowed to make
$$A,\hskip.3in  \lambda,\hskip.15in {\rm and} \hskip.15in\zeta>0.$$
%############################################
%Comparing this with the Wolfenstein formula, one obtains, from 
%$V_{ub}$  $\rho=cos\delta-\zeta$ and $\eta=sin\delta$. ON the other hand,
%from $V_{td}$ one obtains $(1-\rho)^2+\eta^2=\zeta^2.$ The two 
%results actually contradict each other. This discrepancy can also come
%from their approximations, so the relations can not be taken as accurate ones.
%##############################################
In the following, we will give  a brief discussion of
 the values of the parameters in the matrix. Most experimental data
are extracted from  a recent fit by Ali and London\cite{Ali}.
\vskip.08in
1) From the width of some $d$ decays, one has
$$\lambda_d=0.2298\pm 0.0010$$
and from the Kaon decays, 
$$\lambda_s=0.2205\pm 0.0018$$
There is a 5.2 standard deviation discrepancy  between the two values
of $\lambda$.\footnote{With a previous value of $\lambda_s$\cite{databook}, 
the discrepancy 
was  4.5 standard deviations.} 
The significance of the discrepancy must be refined. A scenario which 
needs non-standard 
physics to explain may emerge (or is already there). 
 Before the discrepancy is clarified, the weighted average
 of the two 
\beq
\lambda=0.2270\pm 0.0011,
\eeq
%11
will be used in accurate calculations.
 $\lambda =0.23$ will be used as a round off number.
\vskip.08in
 
2) The life-time of the b quark is simply proportional to $A^2\lambda^4$, 
from which one obtains
\beq
A=0.81\pm 0.058.
\eeq
%12

3) The $B_d-\bar B_d$ mixing mass $\Delta m=0.464\pm 0.018
$ (ps)$^{-1}$\cite{Gibsons} 
is simply proportional to $A^2
\zeta^2\lambda^6$, which simply provides a value of $\zeta$, 
according to the formula
\beq
2{\rm Re} M_{12} = -{\rm Re}\ \left(\frac{G_F^2}{6\pi^2}f_{B_d}^2B_{B_d}M_{B_d}
\right)
(V_{tb}V_{td}^*)^2.
\eeq
%13
This equation involves a product of unknown parameters $f_{B_d}^2B_{B_d}$.
According to  lattice calculations\cite{databook}\cite{Ali},
$\sqrt{B_B}f_B=140 - 240$ MeV,
one obtains 
\beq
\zeta = 0.73\pm 0.29.
\eeq
%14 
This number has a  relatively large theoretical uncertainty.
\vskip.08in
4) The value of   $|V_{ub}/V_{cb}|= 0.08\pm 0.02$, 
provides simply  a constraint to sin$\delta$ 
$$
\sqrt{(\cos\delta-\zeta)^2+\sin^2\delta}=0.363\pm 0.073.
$$
Substituting the $\zeta$ value in (14) into this equation does not shed 
any light on the values of the CP phase $\delta.$
One simply has
\beq 
|\sin\delta|\le 0.363\pm 0.073.
\eeq
%15
Note that this upper-bound of the CP phase is obtained from a CP independent 
measurement.  
One notices that a small $b$ to
$u$ decay strongly excludes the scenario of a maximal CP violation of
sin$\delta=\pm 1$. Furthermore, there is a potential danger that a small
product of $A^2\zeta$sin$\delta$ $(J<1.1\times 10^{-5}$ to $
4.9 \times 10^{-5})$ may not be able to 
explain the observed CP violation in the $K^0$ system. A direct 
measurement of the CP phase sin$\delta$ is badly needed. Such a 
measurement will give an accurate value of sin$\delta$ and 
will immediately test whether the CKM matrix is solely responsible
for all CP violation phenomena, by checking with the relatively tight 
upper-bound Eq (15). In addition, by obtaining a sin$\delta$ value directly, 
one can use the $|V_{ub}/V_{cb}|$  data as the information source for the
parameter $\zeta$, if the sin$\delta$ value from another source agrees with
the bound in (15). 
\vskip.08in
A quantitative comparison of information
from CP violation in K decays with Eq (15) 
needs an extensive effort, because of the
controversies in $B^K$ as well as in the ``long distance" contribution to
 $M_{12}(K)$. A recent effort is made in  Ref\cite{Ali} and references
therein. The results from Eqs (11, 12, 14, 15) are summarized in Table II.
\vskip.08in   
It is easy to find that the
 expressions for both Re$\epsilon_{_K}$ and Re$\epsilon_{_{B_d}}$ are 
simple with this matrix (Re$\epsilon_{_K}$ is simple already 
in popular CKM matrices).
Indeed, one  has
\ben
\bea{ccccc}
{\rm Re} \epsilon_{_{B_d}}& =& {\rm Im}(\Gamma_{12}(B_d))/2\Delta m_{B_d},&&\\ 
{\rm Re} \epsilon_{_K}   &=&\frac{Im M_{12}^*(K)}
{\Delta m_K\left(\frac{2|M_{12}|}{|\Gamma_{12}|}+\frac{|\Gamma_{12}|}{2|M_{12}|}
\right)}
&\simeq &{\rm Im}M^*_{12}(K)/2\Delta m_K.
\eea
\een
%16
the formula for Re$\epsilon_{_K}$ is accurate to 0.3\%.
It is also found that $\epsilon_{_D}$ is small with this matrix, because
the dominant CKM factor in the mixing mass is
$(V_{cs}V_{us}^*)^2$, which is real and positive. 
Noting that Re$\epsilon_{_K}\simeq +0.16\%$\cite{databook}, the formula for 
Re$\epsilon_{_K}$  requires sin$\delta > 0$, if $B^K > 0$ 
(as indicated by lattice calculations).
\vskip.08in
It is intersting to estimate sin$\delta$ from the measured value of
Re$\epsilon_{_K}$, neglecting 
the long distance contribution to Re$M_{12}(K)$. The value of $B^K$ becomes
unimportant, once $B^K$ is positive. One then obtains 
$${\rm Re}\epsilon_{_K}=\frac{1}
{2}A^2\zeta\lambda^4(1+\frac{\eta_3}{\eta_1}ln\frac{m_t^2}{m_c^2})
\sin\delta$$
where $\eta_3$ and $\eta_1$ are QCD correction factors. Additional 
uncertainties in this formula include the error in QCD corrections and
the ambiguity in the c-quark mass. 
Taking $m_c=1.4$ MeV/c$^2$, one obtains $\zeta\sin\delta= 0.27\pm 0.02$.
Combining this with that from $|V_{ub}/V_{cb}|$, one obtains two solutions:
sin$\delta =0.20\pm 0.04, \ \zeta = 
1.35\pm 0.31$ and sin$\delta = 0.42\pm 0.08,\ 
\zeta = 0.64\pm 0.13.$ Both solutions seem reasonable, 
compared with the bounds obtained before, although some 
values are close to the margins and may subject to a challenge, if the margin
becomes more strigent with better data. 
\vskip.1in 

Using the parameters defined in (10), the CP violating event rate $R$ and
the overlap $\sigma$ for different systems are listed in the following table.\\
 \begin{center}
{\bf  Table I. CP violating event rate $R$ and overlap $\sigma$ for 
different systems. }
\vskip.2in
\begin{tabular}{|c|c|ccccc|} \hline \hline
&&&&&&\\ 
System\hskip.8in  & R \hskip.5in&&&
$|\sigma (overlap)|$&&\\ 
\hline
&&&&&&\\
$K^0-\bar K^0$&$ A^2\zeta\lambda^4\sin\delta$ &
$1.7\times 10^{-3}$ &$\sim $&$A^2\zeta\lambda^4\sin\delta$&&\\
&&&&&&\\
$B_d^0-\bar B_d^0$&$\zeta\lambda^2\sin\delta$ &&& 
$\pi\sin\delta\frac{m_c^2}{m_t^2}$&$ <$&$ 2.2\times 10^{-4}$\\
&&&&&&\\
$B_s^0-\bar B_s^0$&$\zeta\lambda^2\sin\delta$ &&& 
$\pi\zeta\lambda^2\sin\delta\frac{m_c^2}{m_t^2}$&$<$&$1\times 10^{-5}$\\
&&&&&&\\
$D^0-\bar D^0$&$A^2\zeta\lambda^6\sin\delta$ &&& 
$A^2\zeta\lambda^4\sin\delta$&$<$&$1\times 10^{-3}$\\
\hline \hline
\end{tabular}

\end{center}

\noindent In the last column of table I, 
the number in front of the quantity is measured and the
number after the quantity is estimated. 
Here $R$ is the upper-bound branching ratio 
of any CP violating processes
in terms of CKM parameters.
For example, for the Kaon, $R = \Gamma(K_L \rightarrow 2\pi)/\Gamma_{K_S}$, and 
for $B_d$, $R(B_+\rightarrow
\psi K_S)=\Gamma(B_+\rightarrow \psi K_S)/\Gamma_{_{B_d}}.$ 
One notices from Table I that none 
of the neutral mixing systems has a large overlap $\sigma$ parameter. 

\begin{center}
{\bf  Table 	II. Values of CKM Parameters }
\vskip.2in
\begin{tabular}{|c|c|c|c|} \hline \hline
&&&\\
$\lambda$&$A$&$\zeta$&sin$\delta$\\ \hline
&&&\\
0.2270$\pm$0.0011&0.81$\pm$
0.058&0.73$^*\pm$0.29&0\ {\rm to}\ 0.44\\
\hline
\end{tabular}\\
{\small $^*$ \ values that are strongly dependent on theoretical prejudices.}
\end{center}
\vskip.15in

It is worth exploring further the implications of Eq (8), in which 
the measurable
CP asymmetry quantities are exclusively expressed in terms of CP phases in the
 decay amplitudes. As explained before, this formula is valid  when the 
matrix (10) is used  for the $B_d$ system. 
 $\epsilon_{_{B_d}}$ in this case is much smaller than the phases in $A_f$.
One therefore has, for  measurable CP violating quantities 
\beq
{\rm Im}\lambda_{B_d\rightarrow \psi K_S}={\rm Im}(-V_{cb}V_{cd}^*/V_{cb}^*
V_{cd})=\sin2\delta,
\eeq
%17
and
\beq
{\rm Im}\lambda_{B_d\rightarrow \pi \pi}=\frac{2\sin\delta(\cos\delta-\zeta)}
{1+\zeta^2-2\zeta\cos\delta}\ .
\eeq
%18
In obtaining Eq (17), the fact of 
$K_S$ being the mixed state of $K^0$ and $\bar K^0$ is
considered. 
Eq (17) shows that the bench-mark mode is a token of the angle
$\delta$ in the CKM matrix with a two-fold ambiguity, as the sign of 
sin$\delta$ is known. There is a small uncertainty with 
penguin contributions\cite{Xing},
however basically this formula is neat, compared with Eq (13) for
example. We expect this formula
 to be frequently used in the BaBar and Belle
analysis in the near future. A measured value of Eq (18) will provide an
independent information on two values of $\zeta$ if sin$2\delta$ in
(17) is known\footnote{An  interesting point is made by  
Dr. Xing that their matrix in Ref\cite{Xing97}
can express Im$B_d \rightarrow \pi\pi$ as sin$2\delta$ where $\delta$ is the 
CP phase in their matrix.}.  According to Ref\cite{Ali},
$$
0.28<\sin2\delta<0.88.
$$
which implies a sin$\delta$ of
\beq
\sin\delta=0.14-0.51.
\eeq
%19
(Another solution which gives a bigger sin$\delta$ is excluded by a small $b$
to $u$ decay rate.)
This value is consistent with the bound in (15).
Combining Eq (19) with the ratio 
$|V_{ub}/V_{cb}|$, a bound for $\zeta$ is obtained which is parallel to
(15)
$$ 0.48<\zeta<1.32.$$
This bound for $\zeta$ is less restrictive than (14). 
\vskip.08in

If one wants to concentrate on heavy 
flavor physics to acquire all CKM parameters, 
it is suggestive to introduce a CKM matrix which is 
convenient for heavy flavor studies only, 
leaving the Kaon unattended. It has been proved that among 
$\epsilon_{_{K}}, \epsilon_{_{B_d}},$ and $\epsilon_{_{B_s}}
$ only two of them can be made 
small at the same time\cite{Chen},
based on the unitarity of the CKM matrix.  The following
matrix is convenient for $B_d, B_s$ as well as $D^0$ studies:

\ben
V_{HF}=\left(\bea{ccc}
1-\frac{1}{2}\lambda^2&\lambda e^{i\delta}
&A\lambda^3(e^{i\delta}-\zeta)\\
&&\\
-\lambda+A^2\zeta\lambda^5e^{-i\delta}&(1-\frac{1}{2}
\lambda^2)e^{i\delta}&A\lambda^2e^{i\delta}\\
&&\\
A\zeta\lambda^3&-A\lambda^2&1\eea\right).
\een
%20
It should be noted  that the beloved formula Re$\epsilon_K \propto
A^2\zeta\lambda^4\sin\delta$ is not valid with this matrix. This is because
Re$\epsilon_K$ can be very large with this matrix. One must now distinguish
the $\epsilon_{_K}$ defined by Eq(1) and the physically measurable
$\epsilon^0_{_K}
$ (the superscript `$^0$'
is for I-spin zero final state) which equals to
$$
\epsilon_{_K}^0=(2\eta^{+-}+\eta^{00})/3.
$$
The difference between this
matrix and that in Eq(10) is that the phases of the elements in the
second column are all shifted by the same amount. With this matrix, in
addition to Eqs (19, 20), one also finds
\beq
{\rm Im}\lambda_{B_s\rightarrow \psi\phi}=2\zeta\lambda^2\sin\delta.
\eeq
This formula provides another neat means to access the $\zeta$ value, if 
sin$\delta$ is measured. 
 It is worth mentioning that a CKM matrix which surrenders small 
$\epsilon_{_{B_d}}$ and $\epsilon_{_{B_s}}$ is recommended by Fritzsch and 
Xing\cite{Xing97} through a different approach.
%21

From the above study, we conclude that the CKM model has already been 
challenged by the over constrained measurements available to date. 
The confrontations may be sharpened if a suitable choice of the CKM matrix
 is taken for a specific $P-\bar P$ system.
  The choices
considered in this note are based on the 
observation that the overlaps $\sigma$ for all the $P-\bar P$ mixing systems
are small. One therefore can always choose phase conventions to make
the CP phases in the mixing mass and width simultaneously small. The
measurable CP asymmetries are therefore completely expressed by the
CP phases in the decay amplitudes. The results Eq (8) and Eq (17)
are thus obtained, which will be the most useful formulas in the 
BaBar and Belle data analysis, and will be convenient
for further confrontations of the 
new data with the old ones.
The question of whether
these  CKM matrices facilitate
the  discussion of nonstandard models will be studied elsewhere.
\vskip.08in
This work is in part supported by  
 U.S. National Science Foundation HRD. Support from the 
 Center for Applied Radiation Research at Prairie View A\&M University
is greatly acknowledged. He thanks L. Turnball and M. Haire for reading of the
manuscript, and Z.Z. Xing for very useful comments.


\begin{thebibliography}{99}

\bibitem{CKM} N. Cabibbo,
Phys. Rev. Lett. {\bf 10}, 531 (1963);
M. Kobayashi and T. Maskawa,
Prog. Theor. Phys. {\bf 10}, 531 (1973).

\bibitem{Wol} L. Wolfenstein,
Phys.  Rev.  Lett. {\bf 51}, 1945 (1983);
J. P. Silva and L. Wolfenstein, Phys. Rev. D {\bf 55}, 5331 (1997).

\bibitem{Chau} L.-L. Chau  and W.-Y. Keung, Phys. Rev. Lett. {\bf 53},
1802 (1984).

\bibitem{databook} Particle Data Group,
Phys. Rev. D {\bf 55}, R1119 (1996).

\bibitem{Chen} W. Chen and D.D. Wu,
Comm.  Theor. Phys. {\bf 14}, 247 (1990).

\bibitem{Xing97} Z.Z. Xing and H. Fritzsch, CERN-TH-97-201, June 1997
e-print hep-ph/9707215.

\bibitem{Nir} Y. Nir,
To appear in the Proceeding of the 18th International
Symposium on Lepton Photon Interactions, Hamburg, 
Germany, July 28-August 1, 1997.



\bibitem{Sanda} A.B. Carter and A.I. Sanda,
Phys. Rev.  Lett.{\bf 45}, 952 (1980); Phys. Rev. D {\bf 23}, 1567 (1981),
I.I. Bigi and A.I. Sanda, Nucl. Phys. B {\bf 193}, 85 (1981); {\bf 281}, 41
 (1987); B. Kayser,
NSF report, NSF-PT-92-01, 1991.


\bibitem{TWu} T.T. Wu and C.N. Yang,
Phys. Rev. Lett. {\bf 13}, 380 (1964).

\bibitem{Wu81} D.D. Wu,
Phys. Lett. B {\bf 90}, 452 (1980).

\bibitem{Pais} For asymmetry in leptonic decays, see A. Pais and S.B. Treiman,
Phys. Rev. D {
\bf 12}, 2744 (1975);
L.B. Okun, V.I. Zakharov, and B.M. Pontecovo,
 Lett.  Nuovo Cim. {\bf 13}, 218 (1975);
M. Bender and S. Siverman, and A. Soni, Phys.  Rev. Lett. {\bf 43}, 242 (1979).

\bibitem{Inami} T. Inami and C.S. Lim,
Prog. Theor. Phys.{\bf 65}, 297 (1981).

\bibitem{Hagelin} J. Hagelin
Nucl. Phys.{\bf 199}, 123 (1981).


\bibitem{Kabir} P.K. Kabir and D.D. Wu,
unpublished, (1988).

\bibitem{Wu89} D.D. Wu,
Phys. Rev. D {\bf 40}, 806 (1989).


\bibitem{Ali} A. Ali and D. London, DESY Report, DESY {\bf 96-140}, eprint,
hep-ph/9607392; see also
The BaBar Physics Book, in preparation.


\bibitem{Gibsons} L.Gibsons (CLEO Collaboration), 
Invited talk at International 
Conference on High Energy Physics, Warsaw, ICHEP96 (1996).

\bibitem{Xing} For a recent analysis, see A.I. Sanda and Z.Z. Xing,
To appear, 1997.

\end{thebibliography}
\end{document}